\documentclass[a4,semcolor,psfig,english,12pt,epsf,portrait]{article}
\usepackage{amsmath,amsfonts,latexsym,amscd,amssymb,theorem}

\usepackage[T1]{fontenc}
\usepackage{epsfig}
\usepackage{amsmath}
\usepackage{psfrag}

\def\div{\hbox{div}}

\def\R{\hbox{\bf R}}
\def\Z{\hbox{\bf Z}}

\def\d{\displaystyle}

\def\div{\hbox{div}}

\def\D{{\cal D}}

\def\e{\varepsilon}

\def\l{{\lambda}}

\def\<{\langle}
\def\>{\rangle}
\newcommand{\ba}{\begin{eqnarray}}
\newcommand{\ea}{\end{eqnarray}}

\newtheorem{theo}{\bf Theorem}[section]
\newtheorem{lem}[theo]{\bf \textit{Lemma}}
\newtheorem{pro}[theo]{\bf Proposition}
\newtheorem{cor}[theo]{\bf Corollary}

\newtheorem{rem}[theo]{\bf Remark}

\renewcommand{\R}{{\mathbb R}}
\renewcommand{\Z}{{\mathbb Z}}

\begin{document}

\title{\bf Diagonal hyperbolic systems 
with \\ large and monotone data \\ Part I: Global continuous solutions}

\author{
\normalsize\textsc{A. El
  Hajj$^1$, R. Monneau$^2$}}
\vspace{20pt}
\maketitle

\footnotetext[1]{Universit\'e d'Orl\'eans,
Laboratoire MAPMO,
Route de Chartres, 45000 Orl\'eans cedex 2, France}
\footnotetext[2]{\'Ecole Nationale des Ponts et
 Chauss\'ees, CERMICS,
6 et 8 avenue Blaise Pascal, Cit\'e Descartes
 Champs-sur-Marne, 77455 Marne-la-Vall\'ee Cedex 2, France}

 \centerline{\small{\bf{Abstract}}}
 \noindent{\small{In this paper, we study  diagonal hyperbolic
     systems in one space dimension. Based on 
      a new  gradient entropy estimate, we prove the
      global existence of a continuous solution,   
      for large and non-decreasing initial data. We remark that these results cover
       the case of systems which are hyperbolic but not strictly hyperbolic. Physically, this
      kind of diagonal hyperbolic systems appears naturally in the
      modelling of the dynamics of dislocation densities.
 }}

\hfill\break
 \noindent{\small{\bf{AMS Classification: }}} {\small{35L45, 35Q35,
     35Q72, 74H25.}}\hfill\break
  \noindent{\small{\bf{Key words: }}} {\small{Global existence, system of Burgers
      equations, system of non-linear transport equations, non-linear
      hyperbolic system, dynamics of dislocation densities.}}\hfill\break

\vspace{20pt}

\section{Introduction and main result}

\subsection{Setting of the problem}
In this paper we are interested in  continuous solutions to hyperbolic
systems in dimension one. Our work will focus on  
solutions $\d{u(t,x)=(u^i(t,x))_{i=1,\dots, d}}$, where $d$ is an integer,
of hyperbolic systems which are
diagonal, i.e.
\begin{equation}\label{EM:burger}
\partial_t u^i + \l^i(u)\partial_x u^i = 0 \quad \mbox{on} \quad  
(0,+\infty)\times\R,\quad \mbox{for}\quad i=1,\dots,d,   
\end{equation}

\noindent with the initial data:
\begin{equation}\label{EM:initialdata}
u^i(0,x)=u_0^i(x),\qquad \mbox{$x\in\mathbb{R}$, \quad for \quad $i=1,\dots,d$}.
\end{equation}

\noindent  Here $\displaystyle{\partial_t= \frac{\partial}{\partial t}}$ and  
$\displaystyle{\partial_x= \frac{\partial}{\partial x}}$.  Such systems are 
(sometimes) called $(d\times d)$ hyperbolic systems. Our study of system (\ref{EM:burger}) 
is motivated by consideration of models describing the dynamics of dislocation densities  (see the Appendix, Section 
\ref{EM:subsce:model}), which is 

$$\partial_t u^i+ 
\left(\sum_{j=1,\dots,d}A_{ij} u^j\right)\partial_x u^i = 0 \quad\mbox{for} \quad i=1,\dots,d,
$$

\noindent where $(A_{ij})_{i,j=1,\dots,d}$ is a non-negative symmetric matrix. This model 
 can be seen as a special case of system (\ref{EM:burger}).\\

\noindent  For real numbers $\alpha^i \le \beta^i$, let us consider the box
\begin{equation}\label{EM:box}\displaystyle{U= \Pi_{i=1}^d [\alpha^i,\beta^i]}.
\end{equation}

\noindent We consider a given  function  
 $\l=(\l^i)_{i=1,...,d}: U\to \R^d$, which 
 satisfies the following regularity assumption:

$$(H1) \quad \left\{\begin{array}{l}
\mbox{the function $\l\in C^{\infty}(U)$,}\\
\\
\mbox{there exists}\quad M_0 >0  \quad \mbox{such that
       for}\quad i=1,...,d,\\
|\l^i(u)|\le M_0 \quad \mbox{for all}\quad
       u\in  U,\\
\\
\mbox{there exists}\quad M_1 >0  \quad \mbox{such that
       for}\quad i=1,...,d,\\
      
       |\l^i(v)-\l^i(u)| \le M_1 |v-u| \quad \mbox{for all}\quad
       v,u\in U,
\end{array}\right.$$

\noindent  where $\displaystyle{|w|=\sum_{i=1,\dots,d}|w^i|}$, 
for $w=(w^1,\dots,w^d)$. Given any Banach space $(E, \|\cdot\|_{E})$,  
in the rest of the paper we consider the norm on $E^d$:

$$
\|w\|_{E^d}=\sum_{i=1,\dots,d}\|w^i\|_{E}, \quad \mbox{for} \quad w=(w^1,\dots,w^d) \in E^d. 
$$

\noindent We assume, for all $u\in \R^d$, that the matrix

$$\mbox{$(\l^i_{,j}(u))_{i,j=1,...,d}$, \quad where \quad
$\d{\l^i_{,j}=\frac{\partial \l^i}{\partial u^j},}$}$$

\noindent is non-negative in the positive cone, namely

$$(H2) \left|\begin{array}{l}
\mbox{for all}\quad u\in U,\quad
  \mbox{we have}\\
\\
  \displaystyle{\sum_{i,j=1,...,d}\xi_i\xi_j \l^i_{,j}(u) \ge 0\quad \mbox{for
  every}\quad \xi=(\xi_1,...,\xi_d)\in [0,+\infty)^d.}
\end{array}\right.$$

\noindent In (\ref{EM:initialdata}), each component $u_0^i$ of
the initial data $u_0=(u_0^1,\dots,u_0^d)$ is assumed satisfy the
following property:\\

$$(H3)\quad \left.\left\{\begin{array}{l}
\mbox{$\alpha^i \le u_0^i\le \beta^i$,}\\
\\
\mbox{$u_0^i$ is non-decreasing,}\\
\\
\mbox{$\partial_x u_0^i \in L\log L(\R)$,}
      \end{array}\right.\right|\mbox{ for $i=1,\dots,d$,}
$$

\noindent where  $L\log L(\R)$
is the following Zygmund space:

 $$L\log L(\R)=\left\{\mbox{$f\in L^1(\R)$ such that
  $\displaystyle{\int_{\R}|f|\ln\left(e+|f|\right)<+\infty}$}
 \right\}.$$
\noindent  This space is
equipped by the following norm:

$$\|f\|_{L\log
  L(\R)}=\inf\left\{\mu >0:\displaystyle{\int_{\R}}\frac
  {|f|}{\mu}\ln\left(e+\frac {|f|}{\mu}\right)\le 1\right\},$$

\noindent  This norm is due to Luxemburg (see Adams \cite[(13), Page
234]{Adams}).\\

\noindent Our purpose is to show the
existence of a continuous solution $u=(u^1,\dots,u^d)$ such that, for $i=1,\dots,d$, the function $u^i(t,\cdot)$ 
 satisfies  $(H3)$ for all time.

\subsection{Main result}

\noindent It is well-known that for the classical scalar Burgers equation $\d{\partial_t u+ \partial_x \left(\frac {u^2}{2}\right)=0}$, 
the solution stays continuous when the initial data is Lipschitz-continuous and
non-decreasing. We want somehow to generalize this result to the case of diagonal
hyperbolic  systems. In particular,  we say that a function $u_0=(u^1_0,\dots,u^d_0)$ is non-decreasing if
 each component $u_0^i$ is non-decreasing for  $i=1,\dots,d$.

\begin{theo}\textbf{(Global existence of a non-decreasing solution)}\label{EM:th1}\\
Assume $(H1)$, $(H2)$ and $(H3)$. Then, there exists a function $u$ which satisfies for all
$T>0$:\\

\noindent i) {\bf Existence of a weak solution:}

\noindent The function $u$  is solution of (\ref{EM:burger})-(\ref{EM:initialdata}), where 
$$\mbox{$u\in [L^{\infty}((0,+\infty)\times\R)]^d \cap [C([0,+\infty);L\log
L(\R))]^d$ and  $\partial_x u \in [L^{\infty}((0,T);L\log
L(\R))]^d$},$$

\noindent  such that for a.e. $t\in [0,T)$ the function $u(t,\cdot)$ is
non-decreasing in $x$  and  satisfies the following $L^{\infty}$ estimate:

\begin{equation}\label{EM:max_pri}
\|u^i(t,\cdot)\|_{L^{\infty}(\R)}\le \|u^i_0\|_{L^{\infty}(\R)}, \quad \mbox{for
 \quad $i=1,\dots,d$},
\end{equation}
\noindent and the  gradient entropy estimate:
\begin{equation}\begin{array}{ll}\label{EM:entropy}
\displaystyle{
\hspace{-1cm}\int_{\R}\sum_{i=1,\dots,d}f\left(\partial_x
  u^i(t,x)\right)dx}+ 
\int_{0}^t\int_{\R} \displaystyle{\sum_{i,j=1,\dots,d}\l^i_{,j}(u)\partial_x
u^i(s,x)\partial_x u^j(s,x)\;dx\;ds}\le C_1,
\end{array}\end{equation}

\noindent where 
\begin{equation}\label{EM:f} 0 \le f(x)=\left\{\begin{array}{ll}
x\ln(x)+\frac{1}{e} & \quad \mbox{if}\quad x \ge 1/e,\\
0 & \quad \mbox{if}\quad 0\le x \le 1/e,\\
\end{array}\right.\end{equation}

\noindent and $C_1\left(T,d,M_1,\|u_0\|_{[L^{\infty}(\R)]^d},\|\partial_x u_0\|_{[L\log L
  (\R)]^d}\right)$.\\ 

\noindent ii) {\bf  Continuity of the solution:}

\noindent The solution $u$ constructed in (i) belongs to $\left[C([0,+\infty)\times
\R) \right]^d$ and  there exists 
a modulus  of continuity $\omega(\delta,h)$,  such
that for all $\delta,h\ge 0$ and  all $(t,x)\in(0,T-\delta)\times \R$, we have:

\begin{equation}\label{EM:contu}|u(t+\delta,x+h)-u(t,x)|\le C_2\;\omega(\delta,h)
\;\;\mbox{with}\;\; \displaystyle{\omega(\delta,h)=\frac{1}{\ln(\frac
    {1}{\delta}+1)}+\frac{1}{\ln(\frac {1}{h}+1)}},
\end{equation}
\noindent where $C_2 \left(T,d,M_0, M_1,\|u_0\|_{[L^{\infty}(\R)]^d},\|\partial_x u_0\|_{[L\log L
  (\R)]^d}\right)$.
\end{theo}


\noindent The key point to establish Theorem \ref{EM:th1} is the gradient entropy
 estimate (\ref{EM:entropy}). We first consider the parabolic regularization of the system (\ref{EM:burger}) and we 
 show that the smooth solution admits the $L^{\infty}$ bound (\ref{EM:max_pri}) and the fundamental  gradient entropy inequality
(\ref{EM:entropy}). Then, these {\it a priori} estimates  will allow us to
pass to the limit when the regularization vanishes,  which will provide the
existence of a solution.  Let us mention that a similar  gradient entropy inequality  was
introduced in  Cannone et al. \cite{EC} to prove  the
existence of a solution of a 
two-dimensional system of two coupled transport equations.

\begin{rem} Remark  that assumption $(H2)$ implies that the second term on the left
hand side of (\ref{EM:entropy}) is non-negative. This will imply 
 the $L\log L$ bound on  the gradient  of the solutions. 
\end{rem}

\noindent Up to our knowledge,  the result stated in Theorem \ref{EM:th1} seems new.  
In relation with our result, we can cite the paper of Poupaud \cite{Poupaud}, 
where a result  of existence and uniqueness of Lipschitz solutions is proven for a 
particular quasi-linear hyperbolic system.\\

\noindent Hyperbolic systems (\ref{EM:burger}) in the case $d=2$ are called strictly
 hyperbolic if and only if we have: 

\begin{equation}\label{stricly_d2}\l^1(u^1,u^2)<\l^2(u^1,u^2). 
\end{equation}

\noindent In this case, a result of Lax
\cite{Lax} implies the existence of Lipschitz monotone solutions of 
(\ref{EM:burger})-(\ref{EM:initialdata}). This result was also extended  by 
Serre \cite[Vol II]{Serre12} in the case of $(d\times d)$ 
rich hyperbolic systems (see 
also Subsection \ref{EM:ref} for more related references). Their results are limited to the case of strictly
hyperbolic systems. On the contrary,  in Theorem \ref{EM:th1}, we do not assume that the  hyperbolic system is 
strictly hyperbolic.  See the following remark for a quite detailed example.

\begin{rem}{\bf (Crossing eigenvalues)}\label{EM:cross}\\
 Condition (\ref{stricly_d2}) on the
eigenvalues is not  required in our framework (Theorem \ref{EM:th1}). Here is a  
 simple example of a $(2\times2)$  hyperbolic but not strictly hyperbolic
 system. We consider solution $u=(u^1,u^2)$ of 

\begin{equation}\label{EM:exem}
\left.\left\{\begin{array}{ll}\partial_t u^1+cos(u^2)\partial_x u^1 =0,\\
\\
\partial_t u^2+u^1sin(u^2)\partial_x u^2
=0,\end{array}\right.\right|\quad\mbox{on \quad $(0,+\infty)\times \R$.}
\end{equation}

\noindent Assume:\\ 

\noindent i)  $u^1(-\infty)=1$, $u^1(+\infty)=2$ and $\partial_x u^1\ge 0$,\\

\noindent ii) $u^2(-\infty)=-\frac{\pi}{2}$,
$u^2(+\infty)=\frac{\pi}{2}$ and $\partial_x u^2\ge 0$.\\

\noindent Here the eigenvalues
$\lambda^1(u^1,u^2)=cos(u^2)$ and $\lambda^2(u^1,u^2)=u^1sin(u^2)$
cross each other at the initial time (and indeed for any
time). Nevertheless, we can compute

 $$(\l^i_{,j}(u^1,u^2))_{i,j=1,2}=\left(\begin{array}{ccc}  0& -sin(u^2) \\
                              sin(u^2)& u^1cos(u^2)         
\end{array}\right),$$
\noindent which satisfies $(H2)$ (under assumptions (i) and (ii)). Therefore
Theorem \ref{EM:th1} gives the existence of a solution to (\ref{EM:exem}) with in particular
(i) and (ii). 
\end{rem}

\begin{rem}{\bf (A generalization of Theorem \ref{EM:th1})}\\
In  Theorem \ref{EM:th1} we have considered
 a particular system in order  to simplify the presentation. Our
approach can be easily  extended to the following generalized system:
\begin{equation}\label{EM:bur1}
\partial_t u^i + \l^i(u,x,t)\partial_x u^i = h^i(u,x,t) \quad \mbox{on} \quad  
(0,+\infty)\times\R,\quad \mbox{for}\quad i=1,...,d,   
\end{equation}

\noindent with  the following conditions:\\

\noindent - $\l^i\in W^{1,\infty}(U\times
\R\times[0,+\infty))$ and the matrix $(\l^i_{,j}(u,x,t))_{i,j=1,...,d}$
is positive in the positive cone for all $(u,x,t)\in U\times
\R\times[0,+\infty)$ (i.e. a condition analogous to $(H2)$).\\

\noindent - $h^i\in W^{1,\infty}(U\times
\R\times[0,+\infty))$,  $\partial_x h^i \ge 0$ and $h^i_{,j} \ge 0$ for all $j \neq i$.

\end{rem}

\noindent  Let us remark that our system 
(\ref{EM:burger})-(\ref{EM:initialdata}) does not create shocks because the
solution (given in Theorem \ref{EM:th1}) is continuous. In this situation, it seems
very natural to expect the uniqueness of the solution. Indeed the notion of
entropy solution (in particular designed to deal with the
discontinuities of weak solutions) does not seem so helpful in this
context. Even for such a simple system, we then ask the following:\\

{\it {\bf  Open question:}} {\bf  Is there uniqueness of the continuous solution given in Theorem
\ref{EM:th1} ?}\\

\noindent In a companion paper (El Hajj, Monneau \cite{EM2}), we will provide some partial answers
 to this question.

\subsection{Application to diagonalizable systems}\label{EM:dia}
Let us first consider a smooth function $u=(u^1,\dots,u^d)$, solution of
the following non-conservative hyperbolic system:
\begin{equation}\label{EM:lef}
\left\{\begin{array}{ll}\partial_t u(t,x)+F(u)\partial_x u(t,x)=0,
&u\in U,\quad x\in\R,\quad t\in (0,+\infty),\\
\\
u(x,0)=u_0(x) &x\in\R,\end{array}\right.\end{equation}

\noindent  where the
space of states $U$ is now an open subset of $\R^d$, and for each $u$, $F(u)$ is a
$(d\times d)$-matrix and the  map $F$ is of class $C^1(U)$. The system (\ref{EM:lef})  is said $(d\times d)$ hyperbolic, 
if $F(u)$ has $d$ real eigenvalues and is diagonalizable for any given $u$ on the domain under
consideration. By definition, such a system is said to be diagonalizable, if there exists a smooth transformation
$w = (w^1(u),\dots,w^d(u))$ with non-vanishing Jacobian such that (\ref{EM:lef}) can be
equivalently rewritten (for smooth solutions) as the following  system
$$\partial_t w^i + \lambda^i(w)\partial_x w^i =0 \quad\mbox{for} \quad i=1,\dots,d,$$

\noindent where $\l^i$ are smooth functions of $w$. Such functions 
$w^i$ are called strict $i$-Riemann invariant.

\noindent Our approach can give continuous solutions to the diagonalized system, which 
provided continuous solution to the  original system (\ref{EM:lef}). 

\subsection{A brief review of some related literature}\label{EM:ref}

\noindent For a scalar conservation law, which corresponds to system (\ref{EM:lef}) in the case
$d=1$ where $F(u)= h'(u)$ is the derivative of some flux function $h$, the
global existence and uniqueness of $BV$ solutions has been established by Oleinik
\cite{Ole} in one space dimension. The famous paper of Kruzhkov
\cite{Kru} covers the more general class of $L^{\infty}$ solutions, in
several space dimensions. For an alternative approach  based on the
notion of entropy process solutions, see for instance Eymard et
al. \cite{Eymard}. For a different approach based on a kinetic formulation, see 
also Lions et al. \cite{LBT}. \\

\noindent We now recall some well-known results for 
a class of $(2\times 2)$ strictly hyperbolic systems in one space
dimension. This means that
$F(u)$ has two real, distinct eigenvalues satisfying (\ref{stricly_d2}).  As mentioned above, Lax \cite{Lax} proved the 
existence and  uniqueness of
non-decreasing  and smooth solutions
for diagonalized  $(2\times 2)$  strictly hyperbolic systems. In the case of some 
$(2\times 2)$ strictly hyperbolic systems, DiPerna \cite{DiPerna2,
  DiPerna3}  showed the global existence of a $L^{\infty}$ solution. The
proof of DiPerna relies on a
compensated compactness argument, based on the representation of the
weak limit in terms of Young measures, which must reduce to a Dirac mass
due to the presence of a large family of entropies. This result is
based on an the idea of Tartar \cite{Tartar}.\\

\noindent For general  $(d\times d)$  strictly
hyperbolic systems; i.e. where $F(u)$ has $d$ real,  distinct eigenvalues

\begin{equation}\label{EM:str_hy}\lambda^1(u)< \dots<\lambda^d(u),\end{equation}

\noindent  Bianchini and Bressan proved in a very complete paper \cite{Bressan}, a  striking global
existence and uniqueness result of solutions to system (\ref{EM:lef}),
assuming that the initial data has small total variation. This approach 
is mainly based  on a careful analysis of the vanishing viscosity approximation. An 
existence  result has first been proved by Glimm \cite{Glimm} in the special case of 
conservative equations,   i.e. $F(u)=Dh(u)$ is the Jacobian of some flux
function $h$. Let us mention that an existence result has been also 
obtained by LeFloch and Liu \cite{LEF88,LEF93} in the
non-conservative case. \\

\noindent We can also mention that, our system (\ref{EM:burger}) is related
to other similar  models in dimension $N \ge 1$, such as scalar transport equations based on
vector fields with low regularity. Such equations were for
instance studied by Diperna and Lions in \cite{Dep}. They have proved
the existence (and uniqueness) of a solution (in the renormalized
sense), for given vector fields in
$L^1((0,+\infty);W^{1,1}_{loc}(\R^N))$ whose divergence is in
$L^1((0,+\infty); L^{\infty}(\R^N))$. This study was generalized
by Ambrosio \cite{Amb2004}, who considered vector fields in
$L^1((0,+\infty);BV_{loc}(\R^N))$ with bounded divergence. In the
present paper, we work in dimension $N=1$ and prove the existence
(and some uniqueness results) of solutions of the system
(\ref{EM:burger})-(\ref{EM:initialdata}) with a velocity  vector
field $\l^i(u)$, $i=1,\dots,d$. Here, in Theorem \ref{EM:th1},  
 the divergence of our  vector field is only in $L^{\infty}((0,+\infty),
L\log L(\R))$. In this case we proved the existence result   
thanks to the gradient entropy estimate (\ref{EM:entropy}), which gives a better estimate
on the solution.\\


\noindent Let us also mention that for hyperbolic and symmetric systems
in dimension $N \ge 1$,
 G$\mbox{a}^{\hspace{-0.2cm}\circ}$rding 
has proved in \cite{Garding} a local existence and uniqueness
result in $C([0,T); H^s(\R^N))\cap C^1([0,T); H^{s-1}(\R^N))$,
with  $s> \frac N2+1$ (see also Serre \cite[Vol I, Th 3.6.1]{Serre12}), 
this result being only local in time, even in dimension $N=1$.

\subsection{Organization of the  paper}
This paper is organized as follows: in  Section \ref{reg_pb}, we approximate the
system (\ref{EM:burger}), by adding the viscosity term
 ($\e\displaystyle{\partial_{xx} u^i }$). 
Then we show a global
in time existence for this  approximated system.  
Moreover, we show that these solutions are regular
and non-decreasing in $x$ for all $t >0 $. In Section \ref{apriori}, we 
prove the gradient entropy inequality and some other 
$\e$-uniform {\it a priori} estimates.  In Section \ref{EM:preuv}, we prove the 
main result (Theorem \ref{EM:th1}) passing to the limit as $\e$ goes to $0$.  Finally, in the appendix 
(Section \ref{EM:subsce:model}), we derive a model for the dynamics of dislocation densities.

\section{Local existence of an approximated  system}\label{reg_pb}

The system (\ref{EM:burger}) can be written as:
\begin{equation}\label{EM:burgers}
\partial_t u+ \l(u) \diamond \partial_x u=0,
\end{equation}
where $u:=(u^i)_{1, \dots, d}$, $\l(u)=(\l^i(u))_{1, \dots, d}$
 and $u\diamond v$ is the
``component by component product'' of the two vectors
$u=(u^1,\dots,u^d),\ v=(v^1,\dots,v^d)\in\mathbb{R}^d$. 
This is the vector in $\mathbb{R}^d$ whose 
coordinates are given by $(u\diamond v)^i:=u^i v^i$. 
 We now consider the following  parabolic 
regularization of system (\ref{EM:burgers}), for all $0<\e\le 1$:

\begin{equation}\label{EM:burgersapp}\left\{ \begin{array}{ll}
\partial_t u^\e+ \l(u^\e) \diamond \partial_x u^\e= \e\partial_{xx}u^\e\\\\
u^{\e}(x,0)=u^{\e}_0(x),\;\;\; \mbox{ with \quad $u^{\e}_0(x):=u_0\ast\eta_{\e}(x)$,}
\end{array}\right.
\end{equation}

\noindent  where $\displaystyle{\partial_{xx}=\frac{\partial^2}{\partial x^2}}$ and 
 $\eta_{\e}$  is a mollifier verify, $\eta_{\e}(\cdot)=\frac
{1}{\e}\eta(\frac{\cdot}{\e})$, such that $\eta\in
C^{\infty}_c(\R)$ is a non-negative function satisfying
 $\int_{\R}\eta=1$.

\begin{rem}\label{molli_est}
By classical properties of the  mollifier  $(\eta_{\e})_{\e}$ and the fact that
  $u_0^\e\in [L^{\infty}(\R)]^d$, then $ u_0\in
  [C^{\infty}(\R)]^d\cap [W^{2,\infty}(\R)]^d$. Moreover using the non-negativity of 
  $\eta_{\e}$, the second equation of (\ref{EM:burgersapp}) we get that 

$$\|u^{\e, i}_0\|_{L^{\infty}(\R)}\le \|u^{i}_0\|_{L^{\infty}(\R)}, \quad \mbox{for \quad
$i=1,\dots,d$,}$$

\noindent and $(H3)$ also implies that $u^\e_0$ is non-decreasing. 

\end{rem}

\noindent The following theorem is a global existence result
 for the regularized system (\ref{EM:burgersapp}).

\begin{theo}\label{EM:theo:exip_regu}{\bf (Global existence of non-decreasing
    smooth solutions)}\\
Assume $(H1)$ and that the initial data $u^\e_0$ is non-decreasing and satisfies
$u_0^\e\in [C^{\infty}(\R)]^d\cap [W^{2,\infty}(\R)]^d$. Then  
the system (\ref{EM:burgersapp}), admits a solution
$u^{\e}\in [C^{\infty}([0,+\infty)\times\R)]^d\cap  [W^{2,\infty}((0,+\infty)\times\R)]^d$
such that the function $u^{\e}(t,\cdot)$ is non-decreasing for all  $t>0$. 
Moreover, for all  $t >0$, we have the
{\it a priori} bounds: 

\begin{equation}\label{EM:max_pri_e}
\|u^{\e, i}(t,\cdot)\|_{L^{\infty}(\R)}\le \|u^{\e,i}_0\|_{L^{\infty}(\R)}, \quad \mbox{for \quad 
$i=1,\dots,d$},
\end{equation} 

\begin{equation}\label{EM:L_1}\left\|\partial_x u^{\varepsilon,i}\right\|_{L^{\infty}([0,+\infty);L^1(
    \R))} \le 2\|u_0^{\e,i}\|_{L^{\infty}(\R)},  \quad \mbox{for \quad
$i=1,\dots,d$}. 
\end{equation}

\end{theo}

\noindent  The lines of the proof of this theorem are very
standard (see for instance Cannone et al. \cite{EC} for a similar problem). For
this reason, we skip the details of the proof. 
First of all we remark that the estimate (\ref{EM:max_pri_e})
is a direct application of the
Maximum Principle Theorem for parabolic equations  (see
Gilbarg-Trudinger \cite[Th.3.1]{CI-TH}). The regularity of the solution 
follows from a  bootstrap argument. The monotonicity of the solution is a consequence of the  
 maximum principle for scalar parabolic equations applied to $w^{\e}= \partial_x u^{\varepsilon}$ satisfying 
 
$$
\partial_t w^\e+ \l(u^\e) \diamond \partial_x w^\e+ \partial_x(\l(u^\e)) \diamond  w^\e = \e\partial_{xx}w^\e.
$$ 

\noindent  Since $\partial_x u^{\varepsilon}
\ge 0$  this implies  easily  the second estimate (\ref{EM:L_1}).

\section{$\e$-uniform {\it a priori} estimates }\label{apriori}
In this section, we  show some $\e$-uniform estimates  
on the solutions of system (\ref{EM:burgersapp}).\\

\noindent Before going into the proof of the gradient entropy inequality
defined in (\ref{EM:entropy}), we announce the  main idea to establish  this 
estimate. Now, let us set for $w\ge 0$ the entropy function
$$\bar{f}(w)=w\ln w.$$
For any {\it non-negative} test function $\varphi \in C^1_c([0,+\infty)\times\R)$, let us define the following {\it ``gradient entropy''} with
$w^i:=\partial_x u^i$:
$$\displaystyle{\bar{S}(t)}= \int_{\R}  \varphi(t,\cdot) \left(\sum_{i=1,\dots,d}
\bar{f}(w^i(t,\cdot))\right)\ dx.$$
It is very natural to introduce such quantity $\bar{S}(t)$ which in
the case $\varphi\equiv 1$, appears to be nothing else than the total entropy of the
system of $d$ type of particles of non-negative densities $w^i \ge 0$.
Then after two integration by parts,  it is formally possible to deduce from (\ref{EM:burger}) the
equality in the following  {\it gradient entropy inequality} for all $t\ge 0$\\
\begin{equation}\label{EM:eq:entropie}
\displaystyle{\frac{d \bar{S}(t)}{dt} + \int_{\R}\varphi \left(\sum_{i,j=1,\dots,d}
  \l^i_{,j}w^iw^j\right)\ dx \le R(t),} \quad \quad \mbox{for}\quad  t\ge 0,
\end{equation}
with the rest
$$R(t)= \displaystyle{\int_{\R }  \left\{(\partial_t\varphi)
\left(\sum_{i=1,\dots,d}\bar{f}(w^i)\right)+(\partial_x\varphi)
\left(\sum_{i=1,\dots,d} \l^i
 \bar{f}(w^i)\right)\right\}\ dx,}
$$
where we do not show the dependence on $t$ in the integrals.
We remark in particular that this rest is formally equal to zero if $\varphi\equiv
1$.\\

\noindent To guarantee the existence of continuous solutions when $\e=0$, we will
assume later $(H2)$  which guarantees the non-negativity on the second term of 
the left hand side of inequality (\ref{EM:eq:entropie}).\\

\noindent Coming back to a rigorous statement, 
we will prove the following result.

\begin{pro}\label{EM:lemme:entro}{\bf(Gradient entropy inequality)}\\ 
Assume $(H1)$ and  consider a function $u_0 \in
\left[L^{\infty}(\R)\right]^d$ satisfying $(H3)$. For any $0<\e \le 1$,  we consider the
solution $u^\e$ of the system (\ref{EM:burgersapp}) given in
Theorem \ref{EM:theo:exip_regu} with initial data
$u_0^\e=u_0\ast\eta_{\e}$. Then for any $T>0$, there exists a constant $C\left(T,d,M_1,
\|u_0\|_{[L^{\infty}(\R)]^d},\|\partial_x u_0\|_{[L\log L
  (\R)]^d}\right)$ such that 

\begin{equation}\label{EM:keyest}
S(t)+\displaystyle{\int_{0}^t\int_{\R}\sum_{i,j=1,\dots,d}\l^i_{,j}(u^\e)w^{\e,i}w^{\e,j}}
\le C,\;\;\;\mbox{with}\;\;\; S(t)=\displaystyle{\int_{\R}\sum_{i=1,\dots,d} f(w^{\e,i}(t,\cdot))dx}.
\end{equation}

\noindent where   $f$ is defined in (\ref{EM:f}) and $w^{\e}=(w^{\e,i})_{i=1,\dots,d}=\partial_x u^{\e}$.

\end{pro}

\noindent  For the proof of Proposition \ref{EM:lemme:entro}, we need  the following technical lemma:

\begin{lem}{\bf($L\log L$ estimate)}\label{EM:e(0)}\\
Let $(\eta_\e)_{\e \in (0,1]}$ be a non-negative mollifier satisfying $\int_{\R}\eta_\e=1$, let $f$ be the function
defined in (\ref{EM:f}) and $h\in L^1(\R)$ be a non-negative function. Then\\

\noindent i) $\displaystyle{\int_{\R}}f(h)<+\infty$ if and only if
$h\in L\log L(\R)$. Moreover we have the following estimates: 

\begin{equation}\label{estimationLlogL}\displaystyle{\int_{\R}} f(h) 
\le 1+ \|h\|_{L\log L(\R)}+\|h\|_{L^1(\R)}\ln\left(1+\|h\|_{L\log L(\R)}\right),
\end{equation}

\begin{equation}\label{estimationLlogL1}\displaystyle{\|h\|_{L\log L(\R)} }
\le 1+ \int_{\R}f(h)  +\ln(1+e^2)\|h\|_{L^1(\R)}.
\end{equation}

\noindent ii) If $h\in L\log L(\R)$, then for every $\e \in (0,1]$ the function
$h_{\e}=h\ast\eta_{\e}\in L\log L(\R)$ and  satisfies

$$\|h_{\e}\|_{L\log L(\R)} \le C \|h\|_{L\log L(\R)} \quad\mbox{and}\quad\|h-h_{\e}\|_{L\log L(\R)} \rightarrow 0 \quad\mbox{as}\quad
\e\rightarrow 0,$$

\noindent where $C$ is a universal constant.

\end{lem}
\noindent {\bf Proof of Lemma \ref{EM:e(0)}:}\\
\noindent  The proof of (i) is trivial.  To prove estimate  (\ref{estimationLlogL}), we first remark 
that, for all $h\ge 0$ and $\mu \in (0,1]$, we have 

 $$ \left(h\ln(h)+ \frac 1e \right)1\!\!1_{\{h\ge \frac 1e\}}
 \le h\ln(h+e) \le h\ln(e+\mu h)+ |\ln(\mu)|h. 
 $$

\noindent We apply  this inequality with  $\d{\mu= \frac{1}{\max(1, \|h\|_{L\log L(\R)})}}$
 and integrate, we get
 
$$\begin{array}{ll}\d{ \int_{\R}f(h) }
&\le \d{ \frac {1}{\mu}\int_{\R}\mu h\ln(e+\mu h)+  |\ln(\mu)| \|h\|_{L^1(\R)} }\\
\\ 
&\le \d{ \frac {1}{\mu} +  |\ln(\mu)| \|h\|_{L^1(\R)}, }
\end{array}$$ 

\noindent where we have used the definition of $\|h\|_{L\log L(\R)}$. This gives (\ref{estimationLlogL}) using the fact that
 $ \d{\mu \ge  \frac{1}{1+ \|h\|_{L\log L(\R)}}} $. 

\noindent To prove (\ref{estimationLlogL1}), we remark that,  for
  $h \ge \frac 1e$, we  have $e \le e^2 h$ and 
  
  $$h\ln(e+h) \le h\ln(h)+ h\ln(1+e^2)\le f(h)+ h\ln(1+e^2).$$
  
\noindent However, for  $ 0 \le h \le \frac 1e$, we have in particular
$$h\ln(e+h) \le h \ln(1+e^2) .$$

\noindent Therefore 

$$ \int_{\R}h\ln(e+h) \le \int_{\R}f(h) +  \ln(1+e^2)  \|h\|_{L^1(\R)}. $$

\noindent From the definition of $\|h\|_{L\log L(\R)}$, we deduce in
particular (\ref{estimationLlogL1}).  For the proof of (ii) see Adams
\cite[Th 8.20]{Adams}.  
 
 $\hfill\Box$ 

\noindent {\bf Proof of Proposition \ref{EM:lemme:entro}:}\\
\noindent First we want to check that $S(t)$ is well defined. To this end, we remark that 
if $w \ge 0$, then 

$$0\le f(w)\le \frac 1e  1\!\!1_{ \{ w\ge \frac 1e \}}+ w\ln(1+ w).
$$

\noindent Which gives that 

$$\int_{\R} f(w) 
\le \|w\|_{L^1(\R)}\ln\left(1+\|w\|_{L^{\infty}(\R)}\right)+ 
 \int_{\R} \frac 1e  1\!\!1_{ \{ w\ge \frac 1e \}}
\le \|w\|_{L^1(\R)} \left(1+  \ln\left(1+\|w\|_{L^{\infty}(\R)}\right) \right).
$$

\noindent Now by Theorem \ref{EM:theo:exip_regu}, we have  $\partial_x u^{\e}=w^{\e}\in \left[L^{\infty}((0,+ \infty);L^1(\R))\right]^d \cap
[W^{2,\infty}((0,+\infty)\times\R)]^d$. This implies that $S \in L^{\infty}(0,+\infty)$.  We compute

$$\begin{array}{lll}
\displaystyle{\frac{d}{dt}S(t)}
&=\displaystyle{\int_{\R}\sum_{i=1,\dots,d}
  f'(w^{\e,i})(\partial_t w^{\e,i})},\\\\
&=\displaystyle{\int_{\R}\sum_{i=1,\dots,d}
f'(w^{\e,i})\partial_{x}\left(-\l^i(u^{\e}) w^{\e,i}
+\varepsilon\partial_{x}
w^{\varepsilon,i}\right)},\\\\
&=\overbrace{
  \mathstrut\displaystyle{\int_{\R}}\sum_{i=1,\dots,d} \l^i(u^{\e})
w^{\e,i}f''(w^{\e,i})\partial_{x} w^{\e,i}}^{J_1}
\;\overbrace{
  \mathstrut-\displaystyle{\varepsilon\int_{\R}\sum_{i=1,\dots,d}\left(\partial_x
w^{\e,i}\right)^2f''(w^{\e,i})}}^{J_2}. 
\end{array}$$

\noindent Remark that these computations (and the integration by parts) are justified 
because  on the one hand $w^{\e,i}$, its derivatives and $\l^i$ are bounded, and on the other  
hand  $w^{\e,i}$ is in $L^{\infty}((0,+ \infty);L^1(\R))$. We know that $J_2\le 0$ because $f$ is convex.
 To control $J_1$, we rewrite it under the following form 

$$J_1= \displaystyle{\int_{\R}}\sum_{i=1,\dots,d} \l^i(u^{\e})
g'(w^{\e,i})\partial_{x} w^{\e,i},$$

\noindent where 

$$g(x)=\left\{\begin{array}{ll}
x-\frac {1}{e} & \quad \mbox{if}\quad x \ge 1/e,\\
0 & \quad \mbox{if}\quad 0\le x \le 1/e.\\
\end{array}\right.$$

\noindent Then,  we deduce that 

$$\begin{array}{lll}J_1
&=\displaystyle{\int_{\R}}\sum_{i=1,\dots,d} \l^i(u^{\e})
\partial_{x}(g(w^{\e,i}))\\ 
&=-\displaystyle{\int_{\R}}\sum_{i,j=1,\dots,d} \l^i_{,j}(u^{\e})w^{\e,j}
g(w^{\e,i}),\\
&=\overbrace{
  \mathstrut-\displaystyle{\int_{\R}}\sum_{i,j=1,\dots,d}
  \l^i_{,j}(u^{\e})w^{\e,j}w^{\e,i}}^{J_{11}}
\;\overbrace{
  \mathstrut-\displaystyle{\int_{\R}}\sum_{i,j=1,\dots,d}
\l^i_{,j}(u^{\e})w^{\e,j}(g(w^{\e,i})-w^{\e,i})}^{J_{12}}.
\end{array}$$

\noindent  We use the fact that
$|g(x)-x|\le \frac {1}{e}$ for all $x\ge 0$ and $(H1)$, to deduce that

$$\begin{array}{lll}|J_{12}|
&\d{\le\frac {1}{e} d M_1}\left\|w^{\e}
\right\|_{\left[L^{\infty}((0,+\infty),L^1(\R))\right]^d}\\
\\
&\d{\le\frac {2}{e} d M_1}\|u_0\|_{[L^{\infty}(\R)]^d} := C_0(\|u_0\|_{[L^{\infty}(\R)]^d},d,M_1)
\end{array}$$

\noindent where we have use 

\begin{equation}\label{L1_estimate}
\left\|w^{\e,i}\right\|_{L^{\infty}((0,+\infty),L^1(\R))} 
\le 2\|u_0^i\|_{L^{\infty}(\R)}, \quad \mbox{for} \quad i=1,\dots,d,
\end{equation}

\noindent  which follows from Remark \ref{molli_est} 
and Theorem \ref{EM:theo:exip_regu}.  Finally, we deduce that
 
$$\begin{array}{lll}\displaystyle{\frac{d}{dt}S (t)}
&\le J_{11}+J_{12}+J_{2}\\
&\le J_{11}+C_0
.\end{array}$$

\noindent Integrating in time on $(0,t)$, for $0<t<T$, we get that, there exists a
positive constant 
$C \left(T,d,M_1,\|u_0\|_{[L^{\infty}(\R)]^d},\|\partial_x u_0\|_{[L\log L (\R)]^d} \right)$ which is independent of $\e$ by (\ref{estimationLlogL})
 and Lemma \ref{EM:e(0)} (ii) such that  

$$S(t)+\displaystyle{\int_{0}^t\int_{\R}}\sum_{i,j=1,\dots,d}
  \l^i_{,j}(u^{\e})w^{\e,j}w^{\e,i}\le C_0T+S(0)\le C.$$

$\hfill\Box$

\begin{lem}{\bf($W^{-1,1}$ estimate on the time derivative of the
    solutions)}\label{EM:lem:etem}\\
Assume $(H1)$ and that the  function $u_0 \in
\left[L^{\infty}(\R)\right]^d$ satisfies $(H3)$. Then for any $0<\e \le 1$,  the
solution $u^{\e}$ of the system (\ref{EM:burgersapp}) given in
Theorem \ref{EM:theo:exip_regu} with initial data $u_0^\e=u_0\ast\eta_{\e},$
satisfies the following
$\e$-uniform estimate for all $T >0$:

$$\left\|\partial_t u^{\e} \right\|_{ \left[L^{2}((0,T); W^{-1,1}(\R))\right]^d}\leq
C\|u_0\|_{\left[L^{\infty}(\R)\right]^d}.$$

\noindent where $C=C(T, M_0) > 0$ and  $W^{-1,1}(\R)$ is  the dual of the space $W^{1,\infty}(\R).$

\end{lem}
\noindent {\bf Proof of Lemma \ref{EM:lem:etem}:}\\
\noindent The idea to  bound
$\partial_t u^{\e}$ is simply to use  the available 
bounds on the right hand side of the equation (\ref{EM:burgersapp}).
 We will give a proof by duality. We multiply the equation (\ref{EM:burgersapp}) by
$\phi\in \left[L^{2}((0,T), W^{1,\infty}(\R))\right]^d$ and integrate on
$(0,T)\times \R$,  which gives

$$\displaystyle{
\int_{(0,T)\times\R}\phi\cdot\partial_t u^\e =
\overbrace{
  \mathstrut\e\int_{(0,T)\times\R}\phi\cdot\partial_{xx}^2u^\e}^{I_1}
\;\overbrace{ \mathstrut-\int_{(0,T)\times\R}\phi\cdot\left(\l(u^\e) \diamond \partial_x u^\e \right) }^{I_2}}.
$$
\noindent We integrate by parts the term $I_1$, and obtain:
\begin{equation}\begin{array}{ll}\label{EM:I_1}\displaystyle{|I_1|\le
\left|\int_{(0,T)\times\R}\partial_{x}\phi \cdot \partial_{x}u^\e\right|}
&\le
\displaystyle{\|\partial_{x}\phi\|_{\left [L^{2}((0,T), L^{\infty}(\R))\right]^d}
    \|\partial_{x}u^\e\|_{\left [L^{2}((0,T), L^{1}(\R))\right]^d}},\\
\\
&\displaystyle{\le
  2T^{\frac 12}\|\phi\|_{\left[L^{2}((0,T), W^{1,\infty}(\R))\right]^d}
\|u_0\|_{\left [L^{\infty}(\R)\right]^d}},
\end{array}\end{equation}
\noindent where we have  used  inequality  (\ref{L1_estimate}). Similarly, for the term $I_2$, 
we have:  

\begin{equation}\begin{array}{ll}\label{EM:I_2}|I_2|
&\le\displaystyle{ M_0\|\phi\|_{\left
      [L^{2}((0,T), L^{\infty}(\R))\right]^d}
\|\partial_{x}u^\e\|_{\left [L^{2}((0,T), L^{1}(\R))\right]^d}},\\
\\
&\le \displaystyle{2T^{\frac 12}M_0\|u_0\|_{\left
      [L^{\infty}(\R)\right]^d}\|\phi\|_{\left
[L^{2}((0,T), W^{1,\infty}(\R))\right]^d}}.
\end{array}\end{equation}

\noindent Finally,  collecting (\ref{EM:I_1}) and (\ref{EM:I_2}), we get that
there exists a constant $C=C(T,M_0)$ independent of
$0<\e\le 1$ such that:
$$\displaystyle{\left|
\int_{(0,T)\times\R}\phi \cdot \partial_t u^\e \right|
\le
C\|u_0\|_{\left
      [L^{\infty}(\R)\right]^d}
\|\phi\|_{\left[L^{2}((0,T), W^{1,\infty}(\R))\right]^d}}
$$
which gives the announced result. $\hfill\Box$

\begin{cor}\label{EM:born}{\bf ($\e$-uniform estimates)}\\
 Assume $(H1)$ and that the function $u_0 \in
\left[L^{\infty}(\R)\right]^d$ satisfies $(H3)$. Then for any $0<\e\le 1$, the
solution  $u^\e$ of the system (\ref{EM:burgersapp}) given in
Theorem  \ref{EM:theo:exip_regu} with initial data $u_0^\e=u_0\ast\eta_{\e},$
satisfies the following
$\e$-uniform estimate for all $T>0$:
\begin{equation}\label{uniformest}\|\partial_{x}u^\e\|_{\left [L^{\infty}((0,+\infty), L^{1}(\R))\right]^d}
+\left\|u^\e\right\|_{\left[ L^{\infty}((0,+\infty)\times \R)\right]^d}
+\left\|\partial_t{u}^{\e}\right\|_{\left[ L^{2}((0,T); W^{-1,1}(\R))\right]^d}
\le C,\end{equation}

\noindent where  $C=C(T, M_0, \left\|u_0\right\|_{\left[L^{\infty}(\R)\right]^d})$.

\end{cor}

\noindent This Corollary  is a  straightforward consequence of Remark \ref{molli_est}, 
Theorem \ref{EM:theo:exip_regu}, estimate (\ref{L1_estimate}) and Lemma \ref{EM:lem:etem}.

 \section{Passage to the limit and  proof of Theorem \ref{EM:th1}}\label{EM:preuv}

In this section, we prove that the system
 (\ref{EM:burger})-(\ref{EM:initialdata}) admits solutions $u$ in the
distributional sense.  They are the limits  of $u^{\e}$ given by
Theorem \ref{EM:theo:exip_regu} when  $\e\rightarrow
0$. To do this, we
will justify the passage  to the limit as $\e$ tends to $0$ in the
system (\ref{EM:burgersapp}) by using some 
compactness  tools that are presented in a first subsection. 
\subsection{Preliminary results}
First,  for all   open interval $I$ of $\R$, we denote by 
$$L\log L(I)=\left\{\mbox{$f\in L^1(I)$ such that
  $\displaystyle{\int_{I}|f|\ln\left(e+|f|\right)<+\infty}$}
 \right\}.$$
\begin{lem}\label{EM:simo}{\bf (Simon's Lemma)}\\
\noindent Let $X$, $B$, $Y$ be three  Banach spaces, such that we have the following injections 
$$\mbox{$X\hookrightarrow B$ with compact embedding and $B\hookrightarrow Y$ with continuous
 embedding}.$$ 

\noindent Let $T>0$. If $(u^\e)_\e$ is a sequence such that,

$$\|u^\e\|_{L^{\infty}((0,T); X)}+
\left\|\partial_t u^\e\right\|_{L^q((0,T); Y)}\le C,
$$

\noindent  where $q>1$ and $C$ is a constant independent of $\e$, 
then $(u^\e)_\e$ is relatively compact in $L^p((0,T); B)$ for all $1\le
p<q$.

\end{lem}
\noindent For the proof, see  Simon \cite[Corollary 4, Page 85]{SI87}.\\

\noindent In order to show the existence of a solution to system (\ref{EM:burger}) in
Subsection \ref{EM:preuvf}, we will apply this lemma to each scalar component of $u^\e$ in the particular case
where  $X=W^{1,1}(I)$, $B= L^{1}(I)$ and
$Y=W^{-1,1}(I):=(W^{1,\infty}_0(I))'$.\\

\noindent We denote by $K_{exp}(I)$ the class of all measurable function $u$,
defined on $I$, for which,

$$\displaystyle{\int_{I}\left(e^{|u|}-1\right)<+\infty}.$$
 
\noindent The space  $EXP(I)= \left\{\mu u:\quad \mu\ge 0 \quad \mbox{and} \quad u \in K_{exp}(I)  \right\}$
 the linear hull of
$K_{exp}(I)$. This space is supplemented with the following Luxemburg
norm (see Adams \cite[(13), Page
234]{Adams} ):
 
$$\|u\|_{EXP(I)}=\inf\left\{\lambda>0:\displaystyle{\int_{I}
\left(e^{\frac{|u|}{\lambda}}-1\right)\le 1}\right\}.$$

\noindent Let us recall some useful properties of this space.

\begin{lem}\label{EM:hold}{\bf(Generalized Hölder inequality, Adams 
    \cite[8.11, Page 234]{Adams})}\\
Let $h\in EXP (I)$ and $g\in L\log L(I)$. Then $hg\in L^1(I)$, with 

$$\|hg\|_{L^1(I)}\le 2\|h\|_{EXP(I)}\|g\|_{L\log L(I)}.$$
\end{lem}

\begin{lem}\label{EM:continu}{\bf(Continuity)}\\
Let $T>0$. Assume that $u\in L^{\infty}((0, +\infty)\times \R)$ such that 

$$\|\partial_x u\|_{L^{\infty}((0,T); L\log L(\R))}+
 \|\partial_t u\|_{L^{\infty}((0,T); L\log L(\R))} \le C_2 $$

\noindent  Then that for all $\delta,h\ge 0$ and  all $(t,x)\in(0,T-\delta)\times \R$, we have: 

$$ |u(t+\delta,x+h)-u(t,x)|\le 6 C_2   \left( \frac{1}{\ln(\frac
    {1}{\delta}+1)}+\frac{1}{\ln(\frac {1}{h}+1)} \right).$$

\end{lem}

\noindent {\bf Proof of Lemma \ref{EM:continu}:}\\
For all $h>0$ and $(t,x)\in (0,T)\times \R$, we have:

\begin{equation}\label{continuestam}\begin{array}{ll}|u(t,x+h)-u(t,x)|
&\le \d{\left|\int_x^{x+h} \partial_x u(t,y)dy\right|}\\
\\
& \le 2 \d{\|1\|_{EXP(x,x+h)}\|\partial_x u(t,\cdot)\|_{L\log L(x,x+h)}},\\
\\
& \le 2 \d{\frac{1}{\ln(\frac {1}{h}+1)}\|\partial_x u\|_{L^{\infty}((0,T); L\log
  L(\R))}},\\
\\
& \le \d{2C_2\frac{1}{\ln(\frac {1}{h}+1)}},
\end{array}\end{equation}

\noindent where we have used in the second line the generalized Hölder
inequality (Lemma \ref{EM:hold}). Now, we prove the continuity in time, for all
$\delta >0$ and $(t,x)\in (0,T-\delta)\times \R$, we have:

$$\begin{array}{llll}
&\delta|u(t+\delta,x)-u(t,x)|\\ \\
&=\d{\int_x^{x+\delta}|u(t+\delta,x)-u(t,x)|dy},\\
\\
&\le \overbrace{
  \mathstrut{\d{\int_x^{x+\delta}|u(t+\delta,x)-u(t+\delta,y)|dy}}}^{K_1},
+\overbrace{
  \mathstrut{\d{\int_x^{x+\delta}|u(t+\delta,y)-u(t,y)|dy}}}^{K_2},
+\overbrace{
  \mathstrut{\d{\int_x^{x+\delta}|u(t,y)-u(t,x)|dy}}}^{K_3}.
\end{array}$$

\noindent Similarly, as in the last estimate (\ref{continuestam}), we get that:
 
$$\begin{array}{llll}
K_1+K_3&\le \delta \d{\int_x^{x+\delta}|\partial_x  u(t+\delta,y)|dy},
+\delta \d{\int_x^{x+\delta}|\partial_x  u(t,y)|dy},\\
\\
&\le \d{4 C_2\frac{\delta}{\ln(\frac {1}{\delta}+1)}}.
\end{array}$$

\noindent Now, we use that $\partial_t u$ is bounded in $L^{\infty}((0,T); L\log L(\R))$,
to obtain that:

$$\begin{array}{llll}
K_2
&\le \d{\int_x^{x+\delta}\int_t^{t+\delta}|\partial_t u(s,y)| ds\ dy},\\
\\
&\le \d{2 \delta }
  \d{\|1\|_{EXP(x,x+\delta)}\|\partial_t u\|_{L^{\infty}((0,T); L\log L(\R))}}
\le\d{2 C_2\frac{\delta}{\ln(\frac {1}{\delta}+1)}.}
\end{array}$$

\noindent Collecting the estimates of 
$K_1$, $K_2$ and $K_3$, we get that:  

$$|u(t+\delta,x)-u(t,x)|\le \frac{1}{\delta}(K_1+K_2+K_3)\le
\d{6 C_2\frac{1}{\ln(\frac {1}{\delta}+1)}}.$$
 
\noindent This last inequality joint to (\ref{continuestam}) implies the result. 

$\hfill\Box$


\subsection{Proof of  Theorem \ref{EM:th1}}\label{EM:preuvf}
The authors would like to thank T. Gallouët for fruitful remarks 
that helped to simplify of the proof of Theorem \ref{EM:th1}. Before to
 prove Theorem \ref{EM:th1}, we first prove the 
 following result.

\begin{theo}\label{Passage}{\bf (Passage to the limit)}\\
Assume  that $u^\e$ is a solution of system  (\ref{EM:burgersapp}) given by 
Theorem  \ref{EM:theo:exip_regu}, with initial data $u^\e_0=u_0\ast \eta_\e$ where $u_0$ satisfies $(H3)$. If 
we assume that for all $T>0$, there exists a constant 
$C >0$ independent on $\e$, such that: 

\begin{equation}\label{estimationLlogL_x} 
\|\partial_x u^\e\|_{\left[L^{\infty}((0,T); L\log L(\R))\right]^d} \le C,
\end{equation}

\noindent then up to extract a subsequence, the function $u^\e$  converges, as $\e$ goes to zero, to a function
 $u$ weakly-$\star$ in $\left[L^{\infty}((0,+\infty)\times\R)\right]^d$. Moreover,  $u$ is a solution of 
(\ref{EM:burger})-(\ref{EM:initialdata}), and satisfies 

$$\left \{ \begin{array}{ll} 
\| u \|_{\left[L^{\infty}((0,+\infty)\times \R)\right]^d} \le \| u_0 \|_{\left[L^{\infty}(\R)\right]^d},\\
\\
\|\partial_x u \|_{\left[L^{\infty}((0,T); L\log L(\R))\right]^d} \le C,\\
\\
\|\partial_t u \|_{\left[L^{\infty}((0,T); L\log L(\R))\right]^d} \le M_0 C,
\end{array}\right.$$

\noindent and  $u(t,\cdot)$ is non-decreasing in $x$, for all  $t>0$ and
satisfies 

\begin{equation}\label{L1estimate_x}
\|u^i \|_{L^{\infty}((0,+\infty); L^1(\R))}
 \le 2 \|u_0^i\|_{L^{\infty}(\R)}\quad\mbox{for}\quad i=1,\dots,d.
\end{equation}

\end{theo}

\noindent {\bf Proof of Theorem \ref{Passage}:}\\
 \noindent \underline{{\bf Step 1 ($u$ solution of (\ref{EM:burger}))}}:
 First, we remark 
that by estimate (\ref{uniformest}) we know that for any $T>0$, the
solutions $u^{\e}$ of the system
(\ref{EM:burgersapp}) obtained with
the help of Theorem  \ref{EM:theo:exip_regu},  are $\e$-uniformly bounded in
$\left[L^{\infty}((0,+\infty)\times\R)\right]^d$. Hence,  as $\e$
goes to zero, we  can extract a subsequence   still denoted
 by $u^{\e}$, that converges weakly-$\star$  in
$\left[L^{\infty}((0,+\infty)\times\R)\right]^d$ to some limit  $u$. 
Then we want to  show that $u$ is a solution of 
system (\ref{EM:burger}). Indeed, since the passage to the
limit in the linear terms is trivial in $\left[\D'((0,+\infty)\times\R)\right]^d$, it
suffices to pass to the limit in the non-linear term
$$\l(u^\e)\diamond \partial_x u^\e.$$

\noindent According to estimate (\ref{uniformest})  we know that for all open and
bounded interval $I$ of $\R$ there exists a constant $C$  independent on $\e$
such that:

$$\left\|{u}^{\e}\right\|_{\left[ L^{\infty}((0,T); W^{1, 1}(I))\right]^d}
+\left\|\partial_t{u}^{\e}\right\|_{\left[ L^{2}((0,T); W^{-1,1}(I))\right]^d}
\le C.$$

\noindent From the compactness of $W^{1,1}(I)\hookrightarrow
L^{1}(I)$, we can apply Simon's Lemma (i.e. Lemma \ref{EM:simo}), 
with $X=\left[W^{1,1}(I)\right]^d$, $B=\left[L^{1}(I)\right]^d$
and $Y=\left[W^{-1,1}(I)\right]^d$, which shows in particular that

\begin{equation}\label{EM:Linf}\mbox {$u^\e$ is relatively  compact in
  $\left[L^{1}((0,T)\times I)\right]^d$.}\end{equation}

\noindent  Then, we can see that (up to extract a subsequence)
$$\l(u^\e) \to \l(u)\quad \mbox{a.e.}$$

\noindent Moreover, from 
Lemma \ref{EM:hold}, similarly as in (\ref{continuestam}),  we can get, 
 for all  $t\in (0,T)$  the following estimates:

$$\begin{array}{ll}
 \d{\left|\int_{I} \partial_x u^\e(t,y)dy\right|}
 \le \d{2 C\frac{1}{\ln(\frac {1}{|I|}+1)}},
\end{array}$$

\noindent  where $C$ is given in (\ref{estimationLlogL_x}). 
By the previous estimate and the fact that $\l (u^\e)$ is uniformly bounded in  $\left[L^{\infty}((0,+\infty)\times
  \R)\right]^d$ and converges a.e. to  $\l(u)$, we can apply the
Dunford-Pettis Theorem (see Brezis \cite[Th IV.29]{Bre}) and prove that 

$$\l(u^\e)\diamond \partial_x u^\e \to \l(u)\diamond \partial_x u $$

\noindent weakly in $\left[L^{1}((0,T)\times I)\right]^d$. Because this
is true for any bounded open interval $I$, then we can pass to the limit in (\ref{EM:burgersapp}) and get that,

$$\partial_t{u}+\l(u)\diamond \partial_x u =0\quad \mbox{in}\quad \D'((0,+\infty)\times \R).$$

\noindent \underline{{\bf Step 2 ({\it A priori} bounds)}}:  By weakly-$\star$ convergence 
 and from the fact that  $L^{\infty}((0,T); L\log L(\R))$ is the dual of $L^{1}((0,T); E_{exp}(\R))$ (see Adams 
 \cite{Adams} for the definition of the Banach space $E_{exp}(\R)$), we can check that $u$ 
 satisfies the following estimates: 
 
 $$\left\|\partial_x{u}\right\|_{\left[ L^{\infty}((0,T);L\log
    L(\R))\right]^d}\le \d{\liminf_{\e \to 0} }
\left\|\partial_x{u}^{\e}\right\|_{\left[ L^{\infty}((0,T);L\log
L(\R))\right]^d} \le C ,$$

\begin{equation}\label{Linftyestimate_x}\|u\|_{\left[L^{\infty}((0,+\infty)\times \R)\right]^d}
\d{\le \liminf_{\e \to 0}}  \left\|u^\e\right\|_{\left[
     L^{\infty}((0,+\infty)\times\R)\right]^d} \le
 \left\|u_0\right\|_{\left[L^{\infty}(\R)\right]^d}.\end{equation}
 
\noindent Thanks to these two estimates, we obtain that

$$\begin{array}{ll}
 \|\partial_t u \|_{\left[L^{\infty}((0,T); L\log L(\R))\right]^d} 
&\le \|\l(u)\diamond \partial_x u\|_{\left[L^{\infty}((0,T); L\log L(\R))\right]^d}\\
\\
&\le M_0  \| \partial_x u\|_{\left[L^{\infty}((0,T); L\log L(\R))\right]^d}
\le M_0 C. \end{array}$$ 

\noindent Moreover (\ref{L1estimate_x})  follows from
(\ref{Linftyestimate_x}) and the fact that  $u(t,\cdot)$ is
non-decreasing in $x$ (as it was the case for $u^\e$).

\noindent \underline{{\bf Step 3 (Recovering the initial data)}}:  Now  we prove that the initial conditions
 (\ref{EM:initialdata}) coincides
 with $u(0,\cdot)$. Indeed, by the $\e$-uniformly estimate  given in
 Corollary \ref{EM:born}, we can prove easily that, we have 

$$\|u^\e(t)-u_0^\e \|_{\left[W^{-1,1}(\R)\right]^d}\le
Ct^{\frac 12}.$$

\noindent Then, we get 

$$\begin{array}{ll}
\|u(t)-u_0\|_{\left[W^{-1,1}(\R)\right]^d}
&\le  \|u-u_0\|_{\left[L^\infty((0,t); W^{-1,1}(\R)) \right]^d}\\
\\
&\le \d{\liminf_{\e \to 0}}  
\|u^\e-u_0^\e \|_{\left[L^\infty((0,t); W^{-1,1}(\R)) \right]^d}\le Ct^{\frac 12}, 
\end{array}$$

\noindent where we have used the weakly-$\star$ convergence in $L^\infty((0,t);
W^{-1,1}(\R))$ in the second line. This proves that  $u(0,\cdot)=u_0$ in
$\left[\D'(\R)\right]^d$.

$\hfill\Box$

\noindent {\bf Proof of Theorem \ref{EM:th1}:}\\
\noindent \underline{{\bf Step 1 (Existence)}}:  Remark that by
assumption $(H2)$ and estimate (\ref{EM:keyest}), we 
deduce from
  (\ref{estimationLlogL1}) joint to (\ref{L1_estimate}) that, the solution $u^\e$ given  by  
  Corollary \ref{EM:born} satisfies the following estimate:

 \begin{equation}\label{estimation_ux} \|\partial_x u^\e\|_{\left[L^{\infty}((0,T); L\log L(\R))\right]^d} \le C, \end{equation}

\noindent where   $C=C\left(T,d,M_1, \|u_0\|_{[L^{\infty}(\R)]^d},\|\partial_x u_0\|_{[L\log L(\R)]^d}\right)$. Now, we apply  
Theorem \ref{Passage} to prove that,  up to extract a subsequence, the
function $u^\e$  converges, as $\e$ goes to zero, 
to a function
 $u$ weakly-$\star$ in $\left[L^{\infty}((0,+\infty)\times\R)\right]^d$,
 with  $u$ is beeing solution to 
(\ref{EM:burger})-(\ref{EM:initialdata}). 
 Moreover, from Lemma \ref{EM:continu}, we deduce that the function $u$
 satisfies the continuity estimate 
(\ref{EM:contu}). \\

\noindent \underline{{\bf Step 2 (Justification of (\ref{EM:entropy}))}}:  Let 

$$\left\{\begin{array}{ll}
\Gamma_{ij}(u^\e)= \frac 12\left(\l^i_{,j}(u^\e) + \l^j_{,i}(u^\e) \right), \quad \mbox{for} \quad i,j=1,\dots,d,\\
\\
w^\e= \partial_x u^\e. 
\end{array}\right.$$

\noindent  For a general matrix $\Gamma$, where  $^t \Gamma=\Gamma \ge 0 $, let us introduce the square root $B=\sqrt{\Gamma}$ 
of $\Gamma$, uniquely defined by

$$ ^t B= B \ge 0 \quad \mbox{and}\quad B^2= \Gamma. 
$$

\noindent Remark that for non-negative  symmetric matrices,  the map  $\Gamma\longmapsto \sqrt{\Gamma}$
is continuous. Then we can rewrite 

$$\int_0^t \int_{\R} \sum_{i,j=1,\dots,d} \l^i_{,j}(u^\e) w^{\e,i} w^{\e,j}= 
\int_0^t \int_{\R} \left|\sqrt{\Gamma(u^\e)} w^\e\right|^2 \le C,
$$

\noindent where $C$ is given  in (\ref{EM:keyest}). Therefore 
$$\sqrt{\Gamma(u^\e)} w^\e \to q  \quad \mbox{weakly  in
$\left[L^{2}((0,t)\times\R)\right]^d$.}
$$

\noindent   Applying the same argument as in Step 1, of the proof of Theorem \ref{Passage}, 
  for the convergence of $\l(u^\e )\diamond \partial_x u^\e $, we see that 
  
$$\sqrt{\Gamma(u^\e)} \partial_x  u^\e \to \sqrt{\Gamma(u)} \partial_x  u=q  \quad \mbox{weakly  in
$\left[L^{1}((0,t)\times\R)\right]^d$.}
$$

\noindent Therefore, using  the weakly convergence in $L^{2}((0,t)\times\R)$, we get

\begin{equation}\label{estimate_passe}
\int_0^t \int_{\R} \sum_{i,j=1,\dots,d} \l^i_{,j}(u) \partial_x  u^{i}
\partial_x  u^{j} 
=\d{\int_0^t \int_{\R} q^2 }
\le \liminf_{\e \to 0} \int_0^t \int_{\R} \left|\sqrt{\Gamma(u^\e)}\partial_x u^\e\right|^2 
 \le C. 
\end{equation}

\noindent Remark  also that for $w^i =\partial_x  u^{i} $, we have 

{\small $$\begin{array}{ll} \d{\sup_{0\le t \le T} \int_{\R} f(w^i) }
&\le \d{ 1+ \|w^i\|_{L^{\infty}((0,T); L\log
    L(\R))}+\|w^i\|_{L^{\infty}((0,T);  L^1(\R))}\ln\left(1+
\|w^i\|_{L^{\infty}((0,T); L\log L(\R))}\right)}\\
\\
&\le \d{ 1+ \|w^i\|_{L^{\infty}((0,T); L\log
    L(\R))}+2\|u^i_0\|_{L^\infty(\R)}\ln\left(1+
\|w^i\|_{L^{\infty}((0,T); L\log L(\R))}\right):= g[w^{i}]}\\
\\
&\le \d{\liminf_{\e \to 0} g[w^{\e,i}]}\\
\\
&\le \d{ 1+C+2\|u^i_0\|_{L^\infty(\R)}\ln(1+C):=C'},  
\end{array}$$}

\noindent where in the first line we have used (\ref{estimationLlogL}), 
in the second line we have used (\ref{L1estimate_x}), in the third line
we have used  the weakly-$\star$ convergence of $w^{\e,i}$ towards $w^i$ in $L^{\infty}((0,T); L\log L(\R))$ and in the 
 fourth line, we have used  (\ref{estimation_ux}).  Putting this result together with (\ref{estimate_passe}), 
 we get (\ref{EM:entropy}) with $C_1=C+C'$.

$\hfill\Box$

\section{Appendix: Example of the dynamics of dislocation densities}\label{EM:subsce:model}
In this section, we present a model describing the dynamics of
dislocation densities. We refer to Hirth et al. \cite{HL92} for a physical
presentation of dislocations which are (moving) defects in crystals.
Even if the problem is naturally a three-dimensional problem, we will
first assume that the  geometry of the problem is invariant by
translations in the $x_3$-direction. This reduces the problem to the
study of dislocations densities defined on the plane $(x_1,x_2)$ and 
moving in a given direction ${b}$ belonging to the plane
$(x_1,x_2)$ (which is called the ``Burger's vector'').\\

\noindent In Subsection \ref{mod_2D}, we present the 2D-model
with multi-slip directions. In the particular geometry where the dislocations densities
only depend  on the  variable $x=x_1+x_2$, this two-dimensional model reduces to
 a one-dimensional model which is presented in Subsection \ref{EM:mod_1D}.  Finally in 
 Subsection \ref{EM:mod_1D_np}, we explain how to
recover equation (\ref{EM:burger}) as a model for dislocation dynamics with 
$$\d{\l^i(u)=\sum_{j=1,\dots,d}A_{ij}u^j}$$ 
\noindent for some particular
non-negative and symmetric matrix $A$.

\subsection{The 2D-model}\label{mod_2D}

We now present in details the  two-dimensional model. We denote by ${
  X}$ the vector ${ X} = (x_1,x_2) \in \R^2$. We consider a crystal filling
the whole space $\R^2$ and its displacement
$v=(v_1,v_2):\R^2\rightarrow\R^2$, where we have not yet introduced the time
dependence.\\ 

\noindent We introduce the total strain  $\varepsilon(v)=(\varepsilon_{ij}(v))_{i,j=1,2}$ which is a symmetric 
matrix defined by

$$\varepsilon_{ij}(v)=\frac{1}{2}\left(\frac{\partial v_i}{\partial x_j} + 
   \frac{\partial v_j}{\partial x_i}\right).$$

\noindent The total strain can be spitted in two parts:

$$
\varepsilon_{ij}(v) = \e^e_{ij} + \e^p_{ij}\quad \mbox{with} \quad \varepsilon^p=\sum_{k=1,\dots, d}\varepsilon^{0,k} u^k,
$$

\noindent where $\e^e_{ij}$ is the elastic strain and $\e^p_{ij}$  is the plastic strain.
 The scalar function $u^k$  is the
plastic displacement associated to the $k$-th slip system whose 
matrix $\varepsilon^{0,k}_{ij}$ is defined by

$$\varepsilon^{0,k}_{ij}=\frac{1}{2}\left({b}^k_{i}{n}^{k}_{j}
  +{n}^{k}_{i}{b}^k_{j}\right),$$

\noindent where $(b^{k}, n^{k})$ is a family of vectors in $\R^2$, such that ${n}^{k}$ is a unit  vector orthogonal 
to the Burger's vector  ${b}^k$ (see Hirth et al. \cite{HL92} for the definition of the Burger's
 vector of a dislocation)\\

\noindent To simplify the presentation, we assume the simplest possible
periodicity property of the unknowns.\\

{\it \noindent \underline{Assumption $(H)$}:

\noindent i) The function $v$ is $\Z^2$-periodic with $\d{\int_{(0,1)^2}v\;
  d{ X}=0}.$ \\

\noindent ii)  For each $k=1,\dots,d$, there exists $L^k\in\R
^2$ such that $u^k(X)-L^k\cdot{ X}$ is a $\Z^2$-periodic.\\

\noindent iii) The integer $d$ is even with $d=2N$ and  we have for $k=1,\dots,N$:

$$L^{k+N}=L^k,\;\; {n}^{k+N}={n}^{k},\;\;  {b}^{k+N}=-{b}^{k},
\;\;\varepsilon^{0,k+N}=-\varepsilon^{0,k}.$$

\noindent iv) We 
denote by ${\tau}^k \in \R^2$ a unit vector parallel to
${b}^{k}$ such that ${\tau}^{k+N}={\tau}^{k}$. We require that $L^k$ is
chosen such  ${\tau}^{k}\cdot L^k \ge 0$.}\\

\noindent Remark  in particular that the plastic strain $\e^p_{ij}$ is $\Z^2$-periodic 
as a consequence of Assumption $(H)$.  The stress matrix  is then given by 

$$
\sigma_{ij} = \sum_{k,l=1,2} \Lambda_{ijkl}
\varepsilon^e_{kl}\quad\mbox{for}\quad i,j=1,2,
$$

\noindent where $\d{\Lambda=\left(\Lambda_{ijkl}\right)_{i,j,k,l=1,2}}$,
are the constant  elastic 
 coefficients of the material, satisfying for some constant $m>0$:
\begin{equation}\label{EM:coercivite}
\sum_{i,j,k,l=1,2} \Lambda_{ijkl}\varepsilon_{ij}\varepsilon_{kl}\geq m\sum_{i,j=1,2}\varepsilon_{ij}^2,
\end{equation}
for all symmetric matrices $\varepsilon=\left(\varepsilon_{ij}\right)_{ij}$,
{\it i.e.} such that $\varepsilon_{ij}=\varepsilon_{ji}.$\\

\noindent Then the stress is assumed to satisfy the equation of elasticity

$$\displaystyle{\sum_{j=1,2}\frac{\partial 
\sigma_{ij}}{\partial x_j}} =0 \quad\mbox{for} \quad i=1,2.$$

\noindent On the other hand the plastic displacement  $u^k$ is assumed to satisfy the
 following transport equation 
 
 $$ \displaystyle{\partial_t u^k} 
=c^k\displaystyle{{\tau}^k.\nabla u^k}  
\quad \mbox{with}
\quad c^k=\displaystyle{\sum_{i,j= 1,2 }\sigma_{ij}}\varepsilon_{ij}^{0,k}.
$$

\noindent This equation can be interpreted, saying that 

\begin{equation}\label{EM:posi0}
\theta^k= {\tau}^k.\nabla u^k \ge 0,
\end{equation}

\noindent is the density of edge dislocations associated to the  Burger's vector $b^k$ 
moving in the direction ${\tau}^k$ at the velocity $c^k$. Here $c^k$ is also 
called the resolved Peach-Koehler force in the physical literature. In particular, 
we see that the dislocation density $\theta^k$ satisfies the following conservation law

 $$ \displaystyle{\partial_t \theta^k} 
= \div(c^k {\tau}^k\theta^k).
$$

\noindent Finally, for $k=1,\dots,d$,  the functions $u^k$ and $v$ are
then assumed to depend on $(t,{ X})\in (0,+\infty)\times \R^2$ and to be solutions of the coupled
system (see  Yefimov \cite[ch. 5.]{Yef1} and Yefimov, Van der Giessen \cite{Yef}):

\begin{center}\begin{equation}\label{EM:coord_eq:elasdis}\left\{\begin{array} {lll}
\displaystyle{\sum_{j=1,2}\frac{\partial \sigma_{ij}}{\partial x_j}} =
0 \hspace{3.5cm}\mbox{ on $(0,+\infty)\times \R^2$, \hspace{0.3cm}\quad for $i=1,2$},\\

\left.\begin{array} {lll}
\sigma_{ij} = \displaystyle{\sum_{k,l=1,2}} \Lambda_{ijkl}\left
  (\varepsilon_{kl}(v)- \displaystyle{\sum_{k=1,\dots,
    d}\varepsilon^{0,k}_{ij}u^k}   \right)& \mbox{ on $
   (0,+\infty)\times \R^2$},\\

\varepsilon_{ij}(v) =\d{\frac{1}{2}\left(\frac{\partial v_i}{\partial
    x_j}+
\frac{\partial v_j}{\partial x_i}\right)}& \mbox{ on $
   (0,+\infty)\times \R^2$},
\end{array}\right|\mbox{\quad for $i,j=1,2$} \\
\\
\displaystyle{\partial_t u^k} 
=\left( \displaystyle{\sum_{i,j\in\{1,2\}}
    \sigma_{ij}}\varepsilon_{ij}^{0,k}\right)\displaystyle{{\tau}^k.\nabla
u^k}  \hspace{0.3cm}\mbox{ on $
   (0,+\infty)\times \R^2$, \hspace{0.3cm}\quad for $k=1,\dots,d$},
\end{array} \right.
\end{equation}\end{center}

\noindent where $\Lambda_{ijkl}$,  $\varepsilon^{0,k}_{ij}$ are fixed parameters 
previously introduced, and the unknowns of the system are $u=(u^k)_{k=1,\dots,d}$ and the
 displacement $v= (v_1,v_2)$.  Remark also that our
equations are compatible with our periodicity assumptions $(H)$, $(i)$-$(ii)$. \\

\noindent  For a detailed  physical presentation of a model with
multi-slip directions, we  refer to Yefimov, Van der Giessen   
\cite{Yef} and  Yefimov \cite[ch. 5.]{Yef1}  and to  Groma, Balogh 
\cite{Groma} for the case of a model with a single slip direction. See
also Cannone et al. \cite{EC} for a mathematical analysis of the
Groma, Balogh model.
\subsection{Derivation of the 1D-model}\label{EM:mod_1D}

\noindent In this subsection we are interested in a particular geometry where the dislocation  densities
depend only on the  variable $x=x_1+x_2$. This will lead to a
1D-model. More precisely, we make the following:\\

{\it \noindent \underline{Assumption $(H')$}:

\noindent i) The functions $v(t, { X})$ and  $u^k(t, { X})-L^k\cdot{ X}$ 
 depend only on the variable $x=x_1+x_2$.\\

\noindent ii)  For $k=1,\dots,d$, the vector ${\tau}^k=({\tau}^k_1,{\tau}^k_2)$ satisfies 
${\tau}^k_1+{\tau}^k_2>0$ with $\d{\mu^k=\frac {1}{{\tau}^k_1+{\tau}^k_2}}$.\\

\noindent iii) For $k=1,\dots,d$, the vector $L^k=(L_1^k,L_2^k)$ satisfies $L_1^k=L_2^k=l^k$. 
}\\

\noindent For this particular one-dimensional  geometry, we denote by an
abuse of notation the function $v=v(t,x)$ which is $1$-periodic in
$x$.  By assumption $(H')$, $(iii)$, we  can see (again  by an abuse of
notation) that $\d{u=(u^k(t,x))_{k=1,\dots,d}}$ is such that
for $k=1,\dots,d$, $u^k(t,x)-l^k\cdot x$  is  $1$-periodic in $x$.\\

\noindent  Now, we can integrate the equations of elasticity, {\it i.e.}
the first equation of (\ref{EM:coord_eq:elasdis}). Using the
$\Z^2$-periodicity of the  unknowns (see assumption $(H)$, $(i)$-$(ii)$),
and the fact that $\varepsilon^{0,k+N}=-\varepsilon^{0,k}$ (see
assumption $(H)$, $(iii)$), we can easily conclude that  the strain 

\begin{equation}\label{EM:strain}\mbox{$\e^e$ is a linear function of
$\d{(u^j-u^{j+N})_{j=1,\dots,N}}$ \quad and of
$\d{\left(\int_0^1(u^j-u^{j+N})\ dx\right)_{j=1,\dots,N}}$.}
\end{equation}

\noindent This leads to the following Lemma 

\begin{lem}{\bf (Stress for the 1D-model)}\label{EM:1D}\\
Under assumptions $(H)$, $(i)$-$(ii)$-$(iii)$ and $(H')$, $(i)$-$(iii)$
and (\ref{EM:coercivite}), we have 

\begin{equation}\label{EM:stress}-\sigma:\varepsilon^{0,i}=\sum_{j=1,\dots,d}A_{ij}u^j+
\sum_{j=1,\dots,d}Q_{ij}\int_0^1u^j\ dx, \quad \mbox{for $i=1,\dots,N$},
\end{equation}

\noindent where for $i,j=1,\dots,N$

\begin{equation}\label{EM:matrix}\left\{\begin{array}{ll}
&\mbox{$A_{i,j}=A_{j,i}$ \quad and\quad
  $A_{i+N,j}=-A_{i,j}=A_{i,j+N}=-A_{i+N,j+N}$,}\\
\\
&\mbox{$Q_{i,j}=Q_{j,i}$ \quad and \quad
  $Q_{i+N,j}=-Q_{i,j}=Q_{i,j+N}=-Q_{i+N,j+N}$.}
\end{array}\right.\end{equation}

\noindent Moreover the matrix $A$ is non-negative.
\end{lem}

\noindent The proof of Lemma \ref{EM:1D} will be given at the end of this
subsection. \\

\noindent Finally using Lemma \ref{EM:1D}, we can eliminate the stress and  reduce
the problem  to a one-dimensional system of $d$ transport equations only
depending on the function  $u^i$, for
$i=1,\dots,d$. Naturally, from (\ref{EM:stress}) and $(H')$, $(ii)$ this 1D-model
has the following form\\

\noindent \underline{{\bf The 1D-model of the dynamics of dislocation densities}}:
\begin{equation}\label{EM:burger_loc}
\mu^i\partial_t
u^i+\left(\sum_{j=1,\dots,d}A_{ij}u^j+\sum_{j=1,\dots,d}Q_{ij}\int_0^1u^j\ dx\right)
\partial_x u^i=0,\qquad \mbox{on $(0,+\infty)\times \R$, $\;\;$for $i=1,\dots,d$},
\end{equation}

\noindent with from (\ref{EM:posi0})

\begin{equation}\label{EM:croissante}\partial_x u^i\ge 0 \quad \mbox{for
    $i=1,\dots,d$.}
\end{equation}

\noindent Now, we give the proof of Lemma \ref{EM:1D}.\\

\noindent {\bf Proof of Lemma \ref{EM:1D}:}\\
\noindent For the 2D-model, let us consider the elastic energy on the periodic cell
(using the fact that $\e^e$ is $\Z^2$-periodic)

$$E(u,v)=\frac 12\int_{(0,1)^2}\sum_{i,j,k,l=1,2} \Lambda_{ijkl}\e^e_{ij}\e^e_{kl}\ d{ X}
\quad\mbox{with} \quad \e^e_{ij}= \e_{ij}(v)-\sum_{k=1,\dots,d}\varepsilon^{0,k}_{ij}u^k
.$$

\noindent By definition of $\sigma_{ij}$ and $\e^e_{ij}$, we have for
$k=1,\dots,d$

\begin{equation}\label{EM:energy}
\sum_{i,j=1,2}(\sigma_{ij}\varepsilon^{0,k}_{ij})=-E_{u^k}'(u,v).
\end{equation}

\noindent On the other hand using $(H')$, $(i)$-$(iii)$, (with $x=x_1+x_2$) we can check that we can
rewrite  the elastic energy as 
$$E=\frac 12\int_{0}^1 \sum_{i,j,k,l=1,2} \Lambda_{ijkl}\e^e_{ij}\e^e_{kl} dx.$$

\noindent Replacing $\e^e_{ij}$ by its expression (\ref{EM:strain}), we get:

 $$\begin{array}{ll}\d{E}
=
&\d{\frac
  12\int_0^1\sum_{i,j=1,\dots,N}A_{ij}(u^j-u^{j+N})(u^i-u^{i+N}) \ dx }\\
\\
&+
\d{\frac12\sum_{i,j=1,\dots,N}Q_{ij}\left(\int_0^1(u^j-u^{j+N})\ dx\right)
\left(\int_0^1(u^i-u^{i+N})\ dx\right),}\end{array}$$

\noindent for some symmetric matrices $A_{ij}=A_{ji}$,
$Q_{ij}=Q_{ji}$. In particular, joint to (\ref{EM:energy}) this gives
exactly (\ref{EM:stress}) with (\ref{EM:matrix}).\\

\noindent Let us now consider the  functions $w^i=u^i-u^{i+N}$ such
that 

\begin{equation}\label{EM:integ}\d{\int_0^1w^i \ dx}=0 \quad \mbox{for
    i=1,\dots,N.}\end{equation}

\noindent From (\ref{EM:coercivite}),  we deduce that

$$\d{0\le E=\frac
12\int_0^1\sum_{i,j=1,\dots,N}A_{ij}w^iw^j \ dx}.$$

\noindent More precisely, for all
$i=1,\dots,N$ and for all $\bar{w}^i\in \R$, we set 

$$w^i=\left\{\begin{array}{ll}
\bar{w}^i& \quad \mbox{on}\quad [0,\frac 12],\\
-\bar{w}^i & \quad \mbox{on}\quad [\frac 12,1],\\
\end{array}\right.$$

\noindent which satisfies (\ref{EM:integ}). Finally, we obtain that  
$$0\le E=\frac
12\int_0^1\sum_{i,j=1,\dots,N}A_{ij}\bar{w}^i\bar{w}^j\ dx.$$

\noindent Because this is true for every $\bar{w}=(\bar{w}^1,\dots,\bar{w}^N)\in \R^N$, we deduce that $A$
a non-negative matrix. 

$\hfill\Box$

\noindent We refer the reader to El Hajj \cite{EL} and El Hajj, Forcadel
\cite{EF} for a  study in the special case of a single slip direction, i.e. in the case $N=1$.

\subsection{Heuristic derivation of the non-periodic model}\label{EM:mod_1D_np}

\noindent Starting from the model (\ref{EM:burger_loc})-(\ref{EM:croissante})
where for $i=1,\dots,d$, the function $u^i(t,x)-l^i\cdot x$  is  $1$-periodic in $x$, we now want to
rescale the  unknowns to make the periodicity disappear. More precisely,
we have the following Lemma:

\begin{lem}{\bf (Non-periodic model)}\label{EM:non_per}\\
Let $u$ be a solution of  (\ref{EM:burger_loc})-(\ref{EM:croissante}) assuming Lemma
\ref{EM:1D} and $u^i(t,x)-l^i\cdot x$  is  $1$-periodic in $x$. Let

$$u_{\delta}^j(t,x)=u^j(\delta t,\delta x),\quad \mbox{for a small 
$\delta>0$ and  for $j=1,\dots,d$,}
$$

\noindent such that, for all $j=1,\dots,d$

\begin{equation}\label{EM:limit}u_{\delta}^j(0,\cdot)\to \bar{u}^j(0,\cdot),
  \quad\mbox{as}\quad \delta\to 0, \quad\mbox{and}\quad
\bar{u}^j(0,\pm \infty)=\bar{u}^{j+N}(0,\pm \infty).\end{equation}

\noindent Then $\d{\bar{u}=(\bar{u}^j)_{j=1,\dots,d}}$  is formally a
solution of

\begin{equation}\label{EM:limit_p}
\mu^i\partial_t
\bar{u}^{i}+\left(\sum_{j=1,\dots,d}A_{ij}\bar{u}^{j}\right)
\partial_x \bar{u}^{i}=0,\qquad \mbox{on $(0,+\infty)\times \R$},\end{equation}

\noindent where the symmetric matrix $A$ is non-negative and $\partial_x
\bar{u}^{i} \ge 0$ for $i=1,\dots,d$.

\end{lem}

\noindent  We remark that the limit problem (\ref{EM:limit_p}) is of type
(\ref{EM:burger}) when $\mu^i=1$. In particular, there are no reasons to assume that
 this system is strictly hyperbolic in general. Nevertheless, the general case  $\mu^i >0$ 
 can be treated with our approach developed in Theorem \ref{EM:th1}  considering the entropy 
$ \displaystyle{\int_{\R}\sum_{i=1,\dots,d} \mu^i f\left(\partial_x
  \bar{u}^i(t,x)\right)dx}$ instead of $ \displaystyle{\int_{\R}\sum_{i=1,\dots,d} f\left(\partial_x
  \bar{u}^i(t,x)\right)dx}$. \\

\noindent {\bf  Formal proof of Lemma \ref{EM:non_per}:}\\
\noindent Here, we know that  $u_{\delta}^i-\delta l^i\cdot x$ is  $\d{\frac
{1}{\delta}}$-periodic in $x$, and satisfies for $i=1,\dots,d$

\begin{equation}\label{EM:limit_p1}\mu^i\partial_t
u_{\delta}^i+\left(\sum_{j=1,\dots,d}A_{ij}u_{\delta}^j+\delta\sum_{j=1,
\dots,d}Q_{ij}\int_0^{\frac
      {1}{\delta}}u_{\delta}^j\ dx\right)
\partial_x u_{\delta}^i=0,\qquad \mbox{on $(0,+\infty)\times \R$.}
\end{equation}

\noindent To simplify, assume that the initial data
$u_{\delta}(0,\cdot)$ converge to a function $\bar{u}(0,\cdot)$ such
that the function $\partial_x u_{\delta}(0,\cdot)$ inside the interval $\d{ \left(\frac {-1}{2\delta}, \frac {1}{2\delta} \right)}$
 has a support in $(-R,R)$, uniformly
in $\delta$, where $R$ a positive constant. Because of the antisymmetry 
property of the matrix $Q$ (see  assumption (\ref{EM:matrix})), and because of assumption (\ref{EM:limit}), 
we expect heuristically that the velocity in
(\ref{EM:limit_p1}) remains uniformly bounded  as $\delta\to 0$.\\

\noindent Therefore, using the finite propagation speed, we see that,  
 there exists a constant $C$ independent in $\delta$, such that 
$\partial_x u_{\delta}(t,\cdot)$ has a support on $ \d{ (-R-Ct,R+Ct) \subset 
\left(\frac {-1}{2\delta}, \frac {1}{2\delta}\right) }$. Moreover, 
from (\ref{EM:limit}) and the fact that 

 $$\sum_{j=1,\dots,d}Q_{ij}\int_0^{\frac
      {1}{\delta}}u_{\delta}^j\ dx=\sum_{j=1,\dots,N}Q_{ij}\int_0^{\frac
      {1}{\delta}}(u^j_\delta-u^{j+N}_\delta)\ dx,$$
\noindent we deduce that 

$$\sum_{j=1,\dots,d}Q_{ij}\int_0^{\frac
      {1}{\delta}}u_{\delta}^j\ dx,$$

\noindent  remains  bounded uniformly in  $\delta$. Then formally the 
non-local term vanishes and we get for  $i=1,\dots,d$

$$\sum_{j=1,\dots,d}A_{ij}u_{\delta}^j+ \delta\sum_{j=1,
\dots,d}Q_{ij}\int_0^{\frac {1}{\delta}}u_{\delta}^j\ dx \to
\sum_{j=1,\dots,d}A_{ij}\bar{u}^j,\quad\mbox{as}\quad \delta\to 0,$$

\noindent which proves that $\bar{u}$ is solution of (\ref{EM:limit_p}),
with the  non-negative  symmetric matrix $A$. $\hfill\Box$


\section{Acknowledgements }
The first author would like to thank M. Cannone, T. Gallouët  and M. Jazar
 for fruitful remarks that helped in the preparation of the paper. This work was partially
supported  the program ``PPF, programme pluri-formations math\'ematiques
financi\`eres et EDP'', (2006-2010), Marne-la-Vall\'ee University and \'Ecole Nationale
des Ponts et Chauss\'ees and the ANR MICA "Mouvements d'Interfaces, Calcul et Applications" 
(2006-2009).

\bibliographystyle{siam}
\bibliography{biblio}

\def\cprime{$'$}
\begin{thebibliography}{10}

\bibitem{Adams}
{\sc R.~A. Adams}, {\em Sobolev spaces}, Academic Press [A subsidiary of
  Harcourt Brace Jovanovich, Publishers], New York-London, 1975.
\newblock Pure and Applied Mathematics, Vol. 65.

\bibitem{Amb2004}
{\sc L.~Ambrosio}, {\em Transport equation and {C}auchy problem for {$BV$}
  vector fields}, Invent. Math., 158 (2004), pp.~227--260.

\bibitem{Bressan}
{\sc S.~Bianchini and A.~Bressan}, {\em Vanishing viscosity solutions of
  nonlinear hyperbolic systems}, Ann. of Math. (2), 161 (2005), pp.~223--342.

\bibitem{Bre}
{\sc H.~Brezis}, {\em Analyse fonctionnelle}, Collection Math\'ematiques
  Appliqu\'ees pour la Ma\^\i trise. [Collection of Applied Mathematics for the
  Master's Degree], Masson, Paris, 1983.
\newblock Th\'eorie et applications. [Theory and applications].

\bibitem{EC}
{\sc M.~Cannone, A.~El~Hajj, R.~Monneau, and F.~Ribaud}, {\em Global existence
  for a system of non-linear and non-local transport equations describing the
  dynamics of dislocation densities}, to appear in Archive for Rational
  Mechanics and Analysis,  (2007).

\bibitem{DiPerna2}
{\sc R.~J. DiPerna}, {\em Convergence of approximate solutions to conservation
  laws}, Arch. Rational Mech. Anal., 82 (1983), pp.~27--70.

\bibitem{DiPerna3}
{\sc R.~J. DiPerna}, {\em Compensated compactness and general systems of
  conservation laws}, Trans. Amer. Math. Soc., 292 (1985), pp.~383--420.

\bibitem{Dep}
{\sc R.~J. DiPerna and P.-L. Lions}, {\em Ordinary differential equations,
  transport theory and {S}obolev spaces}, Invent. Math., 98 (1989),
  pp.~511--547.

\bibitem{EL}
{\sc A.~El~Hajj}, {\em Well-posedness theory for a nonconservative
  {B}urgers-type system arising in dislocation dynamics}, SIAM J. Math. Anal.,
  39 (2007), pp.~965--986.

\bibitem{EF}
{\sc A.~El~Hajj and N.~Forcadel}, {\em A convergent scheme for a non-local
  coupled system modelling dislocations densities dynamics}, Math. Comp., 77
  (2008), pp.~789--812.

\bibitem{EM2}
{\sc A.~El~Hajj and R.~Monneau}, {\em Diagonal hyperbolic systems with large
  and monotone data. part {II}: Some uniqueness results}, In preparation,
  (2009).

\bibitem{Eymard}
{\sc R.~Eymard, T.~Gallou{\"e}t, and R.~Herbin}, {\em Existence and uniqueness
  of the entropy solution to a nonlinear hyperbolic equation}, Chinese Ann.
  Math. Ser. B, 16 (1995), pp.~1--14.
\newblock A Chinese summary appears in Chinese Ann.\ Math.\ Ser.\ A {\bf 16}
  (1995), no.\ 1, 119.

\bibitem{Garding}
{\sc L.~G{\.a}rding}, {\em Probl\`eme de {C}auchy pour les syst\`emes
  quasi-lin\'eaires d'ordre un strictement hyperboliques}, in Les \'{E}quations
  aux {D}\'eriv\'ees {P}artielles ({P}aris, 1962), \'Editions du Centre
  National de la Recherche Scientifique, Paris, 1963, pp.~33--40.

\bibitem{CI-TH}
{\sc D.~Gilbarg and N.~S. Trudinger}, {\em Elliptic partial differential
  equations of second order}, Classics in Mathematics, Springer-Verlag, Berlin,
  2001.

\bibitem{Glimm}
{\sc J.~Glimm}, {\em Solutions in the large for nonlinear hyperbolic systems of
  equations}, Comm. Pure Appl. Math., 18 (1965), pp.~697--715.

\bibitem{Groma}
{\sc I.~Groma and P.~Balogh}, {\em Investigation of dislocation pattern
  formation in a two-dimensional self-consistent field approximation}, Acta
  Mater, 47 (1999), pp.~3647--3654.

\bibitem{HL92}
{\sc J.~P. Hirth and J.~Lothe}, {\em Theory of dislocations, Second edition},
  Krieger, Malabar, Florida, 1992.

\bibitem{Kru}
{\sc S.~N. Kru{\v{z}}kov}, {\em First order quasilinear equations with several
  independent variables.}, Mat. Sb. (N.S.), 81 (123) (1970), pp.~228--255.

\bibitem{Lax}
{\sc P.~D. Lax}, {\em Hyperbolic systems of conservation laws and the
  mathematical theory of shock waves}, Society for Industrial and Applied
  Mathematics, Philadelphia, Pa., 1973.
\newblock Conference Board of the Mathematical Sciences Regional Conference
  Series in Applied Mathematics, No. 11.

\bibitem{LEF88}
{\sc P.~LeFloch}, {\em Entropy weak solutions to nonlinear hyperbolic systems
  under nonconservative form}, Comm. Partial Differential Equations, 13 (1988),
  pp.~669--727.

\bibitem{LEF93}
{\sc P.~LeFloch and T.-P. Liu}, {\em Existence theory for nonlinear hyperbolic
  systems in nonconservative form}, Forum Math., 5 (1993), pp.~261--280.

\bibitem{LBT}
{\sc P.-L. Lions, B.~Perthame, and E.~Tadmor}, {\em A kinetic formulation of
  multidimensional scalar conservation laws and related equations}, J. Amer.
  Math. Soc., 7 (1994), pp.~169--191.

\bibitem{Ole}
{\sc O.~A. Oleinik}, {\em Discontinuous solutions of non-linear differential
  equations}, Amer. Math. Soc. Transl. (2), 26 (1963), pp.~95--172.

\bibitem{Poupaud}
{\sc F.~Poupaud}, {\em Global smooth solutions of some quasi-linear hyperbolic
  systems with large data}, Ann. Fac. Sci. Toulouse Math. (6), 8 (1999),
  pp.~649--659.

\bibitem{Serre12}
{\sc D.~Serre}, {\em Systems of conservation laws. {I}, {II}}, Cambridge
  University Press, Cambridge, 1999-2000.
\newblock Geometric structures, oscillations, and initial-boundary value
  problems, Translated from the 1996 French original by I. N. Sneddon.

\bibitem{SI87}
{\sc J.~Simon}, {\em Compact sets in the space {$L\sp p(0,T;B)$}}, Ann. Mat.
  Pura Appl. (4), 146 (1987), pp.~65--96.

\bibitem{Tartar}
{\sc L.~Tartar}, {\em Compensated compactness and applications to partial
  differential equations}, in Nonlinear analysis and mechanics: Heriot-Watt
  Symposium, Vol. IV, vol.~39 of Res. Notes in Math., Pitman, Boston, Mass.,
  1979, pp.~136--212.

\bibitem{Yef1}
{\sc S.~Yefimov}, {\em Discrete dislocation and nonlocal crystal plasticity
  modelling}, Netheerlands {I}nstitute for {M}etals {R}esearch, University of
  Groningen, 2004.

\bibitem{Yef}
{\sc S.~Yefimov and E.~Van~der Giessen}, {\em Multiple slip in a
  strain-gradient plasticity model motivated by a statistical-mechanics
  description of dislocations}, International Journal of Solids and Structure,
  42 (2005), pp.~3375--3394.

\end{thebibliography}
\end{document}